\documentclass[final,5p,times,twocolumn]{elsarticle}

\usepackage{amssymb}
\usepackage{bm}% bold math
\usepackage[colorlinks=true,citecolor=magenta,urlcolor=cyan]{hyperref}
\usepackage{xspace}% bold math
\usepackage{amsmath}

 \usepackage{lineno}

\newcounter{bla}

\journal{Computer Physics Communications}
\def\pt{P_{\rm{T}}}

\def\gevc2{\rm{GeV}/\textit{c}^2}

\def\kt{\textbf{\textit{k}}_{\rm T}}

\def\pythia{\textsc{Pythia}\xspace}
\def\3p0{string+${}^3P_0$}
\def\StringSpinner{\texttt{StringSpinner}\xspace}
\def\kpt{\textbf{k}'_{\rm{T}}}
\def\kptkpt{\textbf{k}'^2_{\rm{T}}}

\def\SqT{\textbf{S}_{q\rm{T}}}
\def\SqL{S_{\q\rm{L}}}
\def\SqbarL{S_{\qbar\rm{L}}}
\def\Sq{\textbf{S}_{q}}
\def\Sqbar{\textbf{S}_{\qbar}}
\def\Im{\rm{Im}}
\def\fL{f_{\rm{L}}}
\def\GL{G_{\rm{L}}}
\def\GT{G_{\rm{T}}}
\def\thetaLT{\theta_{\rm{LT}}}

\def\qbar{\bar{q}}

\def\setting#1{{\small\texttt{ #1}}}
\def\settingval#1#2{{\small\texttt{ #1 = } #2}}

\def\GeV{\rm GeV}
\def\q{q}
\def\qp{\q'}
\def\qbar{\bar{q}}
\def\qbarp{\bar{q}'}
\def\fL{f_{\rm L}}
\def\thetaLT{\theta_{\rm LT}}

\def\PTa{P_{1\rm T}}
\def\PTb{P_{2\rm T}}

\def\n{\hat{\textbf{n}}}

\def\X{X}

\def\Xi{\X_i}

\def\pmin{\textbf{p}_-}
\def\pp{\textbf{p}_+}

\def\kbar{\bar{k}}

\def\kVec{\textbf{k}}
\def\kbarVec{\bar{\textbf{k}}}
\def\xq{\hat{\textbf{x}}_q}
\def\yq{\hat{\textbf{y}}_q}
\def\zq{\hat{\textbf{z}}_q}
\def\xqbar{\hat{\textbf{x}}_{\qbar}}
\def\yqbar{\hat{\textbf{y}}_{\qbar}}
\def\zqbar{\hat{\textbf{z}}_{\qbar}}

\def\sigmaq{\boldsymbol{\sigma}_q}
\def\sigmaqp{\boldsymbol{\sigma}_{\qp}}
\def\sigmaqbar{\boldsymbol{\sigma}_{\qbar}}
\def\Iq{I_q}

\def\sigmaXq{\sigma^x_q}
\def\sigmaYq{\sigma^y_q}
\def\sigmaZq{\sigma^z_q}
\def\Iqbar{I_{\qbar}}

\def\sigmaXqbar{\sigma^x_{\qbar}}
\def\sigmaYqbar{\sigma^y_{\qbar}}
\def\sigmaZqbar{\sigma^z_{\qbar}}
\def\Iden{I}

\def\kt{\textbf{k}_{\rm T}}
\def\kpt{\textbf{k}'_{\rm T}}
\def\pt{\textbf{p}_{\rm T}}

\def\kptkpt{\textbf{k}'^2_{\rm T}}

\def\SqT{\textbf{S}_{q\rm T}}
\def\SqbarT{\textbf{S}_{\qbar\rm T}}

\def\ktbar{\bar{\textbf{k}}_{\rm T}}
\def\kptbar{\bar{\textbf{k}}'_{\rm T}}
\def\Pt{\textbf{P}_{\rm T}}
\def\kptbarkptbar{\bar{\textbf{k}}'^2_{\rm T}}

\def\Tr{\rm Tr}

\def\C{\mathcal{C}}
\def\Cqq{\C^{\q\qbar}}
\def\aNN{\hat{a}_{\rm NN}}
\def\CMM{\Cqq_{\rm xx}}
\def\CNN{\Cqq_{\rm yy}}
\def\CMN{\Cqq_{\rm xy}}
\def\CNM{\Cqq_{\rm yx}}
\def\CUL{\Cqq_{\rm 0z}}
\def\CLL{\Cqq_{\rm zz}}
\def\CLU{\Cqq_{\rm z0}}
\def\GZ{\Gamma_Z}
\def\MZ{M_Z}

\def\Pythia{\textsc{Pythia}}

\def\StringSpinner{\texttt{StringSpinner}\xspace}

\def\wh{w_h}
\def\wH{w_H}
\def\VM{\rm VM}
\def\PM{\rm PM}

\def\Im{\rm Im}
\def\Re{\rm Re}
\def\GT{G_{\rm T}}
\def\GL{G_{\rm L}}

\def\AOneTwo{A_{12}}
\def\AOneTwoUL{A_{12}^{UL}}
\def\AOneTwoUC{A_{12}^{UC}}

\def\DOne{D_{1\q}^{h_1}}
\def\DTwo{D_{1\qbar}^{h_2}}

\def\QT{Q_{\rm T}}

\def\HqhOne{H_{1\q}^{\perp\,h_1}}
\def\HqhTwo{H_{1\qbar}^{\perp\,h_2}}

\def\sqrts{\sqrt{s}}

\def\kbar{\bar{k}}
\def\kpbar{\bar{k}'}

\def\i{\rm{i}}

\def\Gammaq{\Gamma}

\def\Trq{\rm Tr_{\q}}

\def\Pbf{\textbf{P}}

\def\CMS{CMS}

\begin{document}

\begin{frontmatter}



%\title{Inclusion of quark-spin effects in \pythia for $e^+e^-$ annihilation \\ with StringSpinner}
\title{StringSpinner 2.0: Enabling quark spin effects in PYTHIA for $e^+e^-$ annihilation}

%A \author[a]{A. Kerbizi\corref{author}}
\author[a,b]{Albi Kerbizi\corref{author}}
\author[a]{Leif L\"onnblad}
%\author[b]{Third Author}

\cortext[author] {Corresponding author.\\\textit{E-mail address:} albi.kerbizi@ts.infn.it}
\address[a]{Department of Physics, Box 118, 221 00 Lund, Sweden}
\address[b]{INFN Sezione di Trieste, Via Valerio 2, 34127 Trieste, Italy}
%\cortext[author] {Corresponding author.\\\textit{E-mail address:} }

\begin{abstract}
The StringSpinner package, which implements quark spin effects in the \Pythia{} event generator using the string+${}^3P_0$ model of hadronization, is extended to handle the process $e^+e^-\rightarrow \q\qbar \rightarrow hadrons$ at leading order. The correlations between the spin states of the $\q\qbar$ pair produced in the reaction are described by a joint spin density matrix. The spin correlations are propagated along the string fragmentation chain by using the rules of the string+${}^3P_0$ model and are implemented for the production of pseudoscalar and vector mesons. The new version of the package can be used to simulate spin effects in $e^+e^-$ annihilation like the Collins and the Artru-Collins asymmetries. It can also be applied to the study of other spin effects in hadronization predicted by the string+${}^3P_0$ model with a user-defined joint spin density matrix of the $\q\qbar$ pair. To showcase the usage of the package the polar angle dependence of the Collins asymmetries for back-to-back pions is studied and compared with the available data.

\end{abstract}


\end{frontmatter}

%\linenumbers

% All CPiP articles must contain the following
% PROGRAM SUMMARY.

{\bf NEW VERSION PROGRAM SUMMARY}
  %Delete as appropriate.

\begin{small}
\noindent
{\em Program Title: StringSpinner}                                          \\
{\em CPC Library link to program files: } (to be added by Technical Editor) \\
{\em Developer's repository link:\\
   \href{https://gitlab.com/albikerbizi/stringspinner.git}{~~~~https://gitlab.com/albikerbizi/stringspinner.git}} \\
{\em Code Ocean capsule:} %(to be added by Technical Editor)
\\
{\em Licensing provisions: GNU GPL v2 or later}  \\
{\em Programming language: C++, Fortran}    \\
%{\em Supplementary material:}                                 \\
  % Fill in if necessary, otherwise leave out.
{\em Journal reference of previous version: }*                  \\
  %Only required for a New Version summary, otherwise leave out.
{\em Does the new version supersede the previous version?: Yes}   \\
  %Only required for a New Version summary, otherwise leave out.
{\em Reasons for the new version: Quark spin effects are enabled for the simulation of $e^+e^-$ annihilation.}\\
  %Only required for a New Version summary, otherwise leave out.
{\em Summary of revisions: The string+${}^3P_0$ model is used to fragment strings stretched between a quark and an antiquark with correlated spin states as produced in $e^+e^-$ annihilation.}\\
  %Only required for a New Version summary, otherwise leave out.
{\em Nature of problem: The $e^+e^-$ annihilation is an important process for the study of the quark spin dependence of hadronization. An event generator including quark spin effects in this process is lacking.}\\
  %Describe the nature of the problem here. \\
{\em Solution method: Extend the previous version of StringSpinner, restricted to deep inelastic scattering, to handle $e^+e^-$ annihilation with quark spin effects.}\\
  %Describe the method solution here.
%{\em Additional comments including restrictions and unusual features (approx. 50-250 words):}\\
  %Provide any additional comments here.
   \\

\end{small}

%\tableofcontents

\section{Introduction} \label{sec:Introduction} 
The quark spin effects were introduced in the string fragmentation part of the \Pythia{} event generator for the simulation of deep inelastic scattering (DIS) at leading order via the \StringSpinner package~\cite{Kerbizi:2021StringSpinner,Kerbizi:2023cde}. To describe the spin effects, \StringSpinner uses the string+${}^3P_0$ model of hadronization~\cite{Kerbizi:2021M20}, which is an extension of the Lund Model (LM) of string fragmentation that includes the quark spin degree of freedom at the amplitude level. The string+${}^3P_0$ model reproduces the Collins effect~\cite{Collins:1992kk} and the dihadron production asymmetry~\cite{Collins:1993kq,Jaffe:1997hf,Bianconi:1999cd} in the fragmentation of transversely polarized quarks, and includes longitudinal spin effects in the fragmentation of longitudinally polarized quarks~\cite{Kerbizi:2018qpp}. The \StringSpinner package thus enables the simulation of the Collins asymmetries~\cite{COMPASS:2014bze,COMPASS:2022jth,HERMES:2010mmo} and the dihadron asymmetries~\cite{COMPASS:2014ysd} in semi-inclusive DIS (SIDIS) with a transversely polarized nucleon target. It can also be used for the simulation of beam spin asymmetries in SIDIS with a longitudinally polarized lepton beam~\cite{Hayward:2021psm}. The production of final state hadrons is presently restricted to pseudoscalar mesons (PMs) and vector mesons (VMs). An extension of the string+${}^3P_0$ model that includes the production and decay of spin-1/2 baryons has been recently proposed in Ref.~\cite{Kerbizi:2025keh}. The implementation of baryon production in \StringSpinner is a major task, which we leave for a separate work.

Another important process for the studies of the quark-spin dependence of hadronization is the annihilation of electrons and positrons to hadrons ($e^+e^-$ annihilation). The $e^+e^-$ pair can annihilate either by the exchange of a virtual photon, $e^+e^-\rightarrow \gamma^*\rightarrow \q\qbar$, or by the exchange of a $Z^0$ boson, $e^+e^-\rightarrow Z^0\rightarrow \q\qbar$. The $\q\qbar$ pair is produced with correlated spin states~\cite{Chen:1994ar}. The correlations between the transverse spins of $q$ and $\qbar$, in particular, allow to probe the transverse-spin dependence of hadronization, namely the Collins FF by the Collins asymmetries~\cite{Boer:2008fr} or the dihadron production asymmetry by the Artru-Collins asymmetries~\cite{Artru:1995zu}. The Collins asymmetries have been measured to be non-vanishing by the BELLE~\cite{Belle:2008fdv,Belle:2019nve} and BABAR~\cite{BaBar:2013jdt,BaBar:2015mcn} experiments at the center of mass energy $\sqrts\simeq 10.58\,\GeV$ and by the BESIII~\cite{BESIII:2015fyw} experiment at $\sqrts=3.65\,\GeV$. The Artru-Collins asymmetries have been measured to be non-vanishing by the BELLE experiment~\cite{Belle:2011cur}.

The joint phenomenological analysis of the asymmetry data from $e^+e^-$ annihilation and SIDIS allows to access both the spin-dependence of hadronization and the partonic transverse-spin structure of the nucleons. An example is the extraction of the transversity parton distribution function (PDF), which describes the transverse polarization of quarks in a transversely polarized nucleon, and the spin-dependent Collins FF responsible for the Collins effect. Given the importance of $e^+e^-$ annihilation for the spin-dependent hadronization studies, it is desirable to build a MC generator capable of reproducing spin effects for such reaction.

In this work we extend the \StringSpinner package of Ref.~\cite{Kerbizi:2023cde} to enable the simulation of $e^+e^-$ annihilation with quark spin-effects in \pythia{}. The new package allows therefore to simulate both $e^+e^-$ annihilation and polarized DIS~\footnote{For the simulation of DIS in the present package the same instructions as in Ref.~\cite{Kerbizi:2023cde} apply and the same results are obtained when using a common setting for the free parameters.}. To implement the spin effects for $e^+e^-$ we begin by setting up the joint spin density matrix $\rho(\q,\qbar)$ of the produced $\q\qbar$ pair. This matrix implements the correlations between the spin states of $q$ and $\qbar$. We consider unpolarized $e^-$ and $e^+$ beams and use the expression of $\rho(\q,\qbar)$ given in Ref.~\cite{Chen:1994ar}, evaluated by accounting for both the $\gamma^*$ and $Z^0$ annihilation channels as well as for the interference between them. The fragmentation of the string stretched between $\q$ and $\qbar$, which have correlated spin states, is described by the string+${}^3P_0$ model using the recipe developed in Ref.~\cite{Kerbizi:2023luv}. Likewise to the previous version of \StringSpinner, the final state parton shower is switched off. The connection between the spin effects in the shower and in string fragmentation has not yet been modeled.

The package described here was used in Ref.~\cite{Kerbizi:2024vpd} for the detailed study of the Collins asymmetries in $e^+e^-$ predicted by the string+${}^3P_0$ model and for the comparison with the available $e^+e^-$ data.

The article is organized as follows. In Sec.~\ref{sec:implementation} we describe the implementation of the spin effects in \pythia{} for $e^+e^-$. In Sec.~\ref{sec:program files} we describe the structure of the StringSpinner package as well as an example of the main program that can be used for the simulation of $e^+e^-$ annihilation. As an example of use of the current \StringSpinner package, we evaluate the polar angle dependence of Collins asymmetries is in Sec.~\ref{sec:results} and compare it to the data by the BABAR experiment~\cite{BaBar:2013jdt}. Finally, in Sec.~\ref{sec:conclusions} we draw the conclusions.

\section{Implementation of spin effects in \pythia for $e^+e^-$}\label{sec:implementation}
\subsection{Setting up the hard reaction $e^+e^-\rightarrow \q\qbar$}
To begin the simulation, we set up the energy $E_-$ of the electron and the energy $E_+$ of the positron, in a frame where the electron momentum $\pmin$ and the positron momentum $\pp$ are in the opposite directions. We let \Pythia{} generate the hard reaction $e^+e^-\rightarrow \gamma^*/Z^0\rightarrow \q\qbar$. It consists in selecting the flavor of the quark $q$ among the kinematically allowed flavors and in constructing the momenta $k$ and $\kbar$ of the produced quark $q$ and antiquark $\qbar$, respectively, according to the corresponding differential cross section. %The annihilation reaction can occur via the exchange of $\gamma^*$ or a $Z^0$ boson. The cross section is thus calculated taking into account the interference between these two channels.

\subsubsection{Helicity frames for $\q$ and $\qbar$}
\begin{figure}[th]
\centering
\begin{minipage}[b]{0.45\textwidth}
%\hspace{-1.0em}
\includegraphics[width=1.0\textwidth]{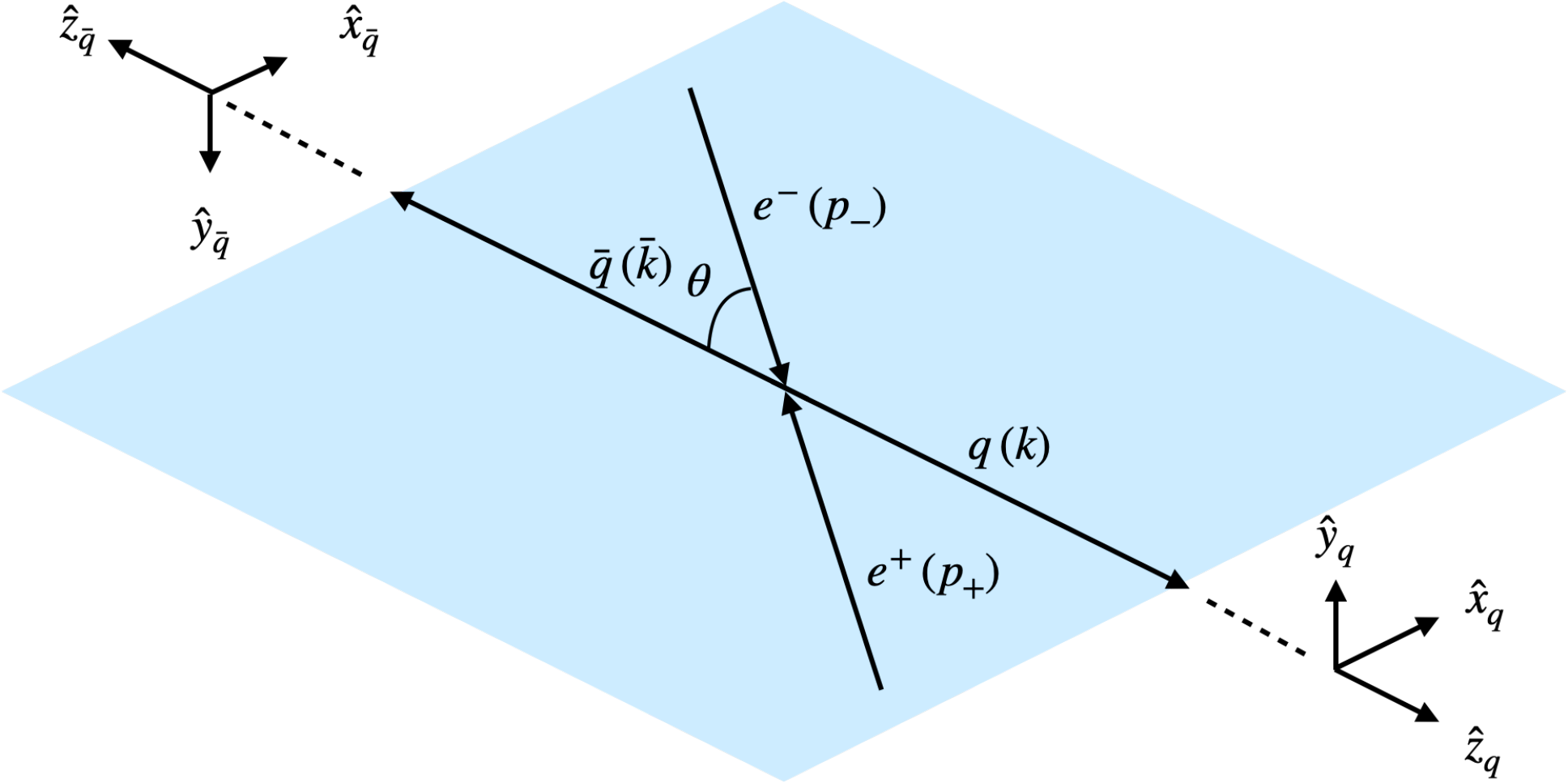}
%\vspace{1.8em}
\end{minipage}
\caption{Kinematics of the $e^+e^-\rightarrow q\qbar$ in the \CMS.}
\label{fig:kinematics}
\end{figure}

The kinematics of the reaction in the center of mass system (\CMS) of $e^+$ and $e^-$ is shown in Fig.~\ref{fig:kinematics}. We indicate with $\theta$ the angle between the momentum $\pmin$ of $e^-$ and the momentum $\kVec$ of $\q$. The momenta of $e^+$ and $\qbar$ are indicated by $\pp=-\pmin$ and $\kbarVec=-\kVec$, respectively. Following Ref.~\cite{Kerbizi:2023luv}, we define the quark helicity frame (QHF) by the set of axes $\lbrace \xq, \yq,\zq\rbrace$, where
\begin{eqnarray}\label{eq:QHF}
     \zq=\kVec/|\kVec|, & \yq=\pmin\times\zq/|\pmin\times \zq |, & \xq= \yq\times \zq.
\end{eqnarray}
 The antiquark helicity frame (AHF) is defined by the axes $\lbrace \xqbar, \yqbar, \zqbar\rbrace$, obtained as in Eq.~(\ref{eq:QHF}) after replacing $\kVec$ with $\kbarVec$. The QHF and AHF are also shown in Fig.~\ref{fig:kinematics}. %In the $e^+e^-$ \CMS, the momenta of $e^-$ and $q$ can be expressed in the QHF as $\pmin=\sqrt{s}\,(-\sin\theta,0,\cos\theta)/2$ and $\kVec=\sqrt{s}\,(0,0,1)/2$, where the electron and the quark masses are neglected.
 In the simulation, we conventionally express all momenta in the QHF.

\subsection{The joint spin density matrix $\rho(\q,\qbar)$}
Once the $\q\qbar$ pair is generated, we set up the joint spin density matrix $\rho(\q,\qbar)$, which implements the correlations between the spin states of $\q$ and $\qbar$. For unpolarized beams, including $\gamma^*/Z^0$ interference and neglecting the quark mass, the joint spin density matrix reads~\cite{Chen:1994ar}
\begin{align}\label{eq:rho}
\nonumber \rho(\q,\qbar)&=\frac{1}{4}\,\bigg[\Iq\otimes \Iqbar +\CLL\, \sigmaZq\otimes \sigmaZqbar + \CUL\,\big(\Iq\otimes\sigmaZqbar-\sigmaZq\otimes \Iqbar \big)\\
\nonumber \hspace{-1.5em} &+\CMM\,\big(\sigmaXq\otimes \sigmaXqbar + \sigmaYq\otimes \sigmaYqbar\big) + \CMN\,\big(\sigmaXq\otimes \sigmaYqbar - \sigmaYq\otimes \sigmaXqbar\big)\bigg]\\
&\equiv \frac{1}{4}\,\C^{\q\qbar}_{\alpha\beta}\,\sigmaq^{\alpha}\otimes \sigmaqbar^{\beta}.
\end{align}
We refer to $\C^{\q\qbar}_{\alpha\beta}$, for $\alpha,\beta=0,x,y,z$, as the correlation coefficients. The matrix $\sigma^{i}_{\q(\qbar)}$ indicates the Pauli matrix along the axis $i=x,y,z$ in the QHF (AHF), while $\sigma^{0}_{\q(\qbar)}=I_{q(\qbar)}$ is the identity matrix in the $q(\qbar)$ spin space. The coefficients are normalized such that $\Cqq_{00}=1$.

Neglecting the quark masses, the non-vanishing coefficients are~\cite{Chen:1994ar}
\begin{align}\label{eq:Cqqbar}
\nonumber \hspace{-2em}   \CUL &= \frac{1}{N_q}\,\bigg\lbrace\,\chi_2(s)\big[v_q\,a_q(v_l^2+a_l^2)\,(1+\cos^2\theta) \\
\nonumber &+ 2\,v_e\,a_e\,(v_q^2-a_q^2)\,\cos\theta\big]\\
\nonumber &-e_q\,\chi_1(s)\big[a_qv_e\,(1+\cos^2\theta)+2\,v_q\,a_e\cos\theta\big]\,\bigg\rbrace\equiv -\,\CLU,\\
\nonumber \CMM &= \frac{1}{N_q}\,\bigg[\frac{e_q^2}{2}+\frac{\chi_2(s)}{s}\,(v_e^2+a_e^2)\,(v_q^2-a_q^2)-e_q\,\chi_1(s)\,v_e\,v_q\bigg]\\
\nonumber &\times \sin^2\theta \equiv \CNN,\\
\CMN &= -\frac{1}{N_q}\,e_q\,\chi_1(s)\,v_e\,a_q\,\frac{\GZ\,\MZ}{s-\MZ^2}\,\sin^2\theta\equiv -N_q^{-1}\,\CNM,
\end{align}
where the function $N_q(s,\cos\theta)$ reads
\begin{align}
\nonumber    N_q(s,\cos\theta)&=\frac{e_q^2}{2}\,(1+\cos^2\theta)+ \frac{\chi_2(s)}{2}\bigg[(1+\cos^2\theta)\,(v_q^2+a_q^2)\\
\nonumber &\times (v_e^2+a_e^2)+8\,v_ea_ev_qa_q\,\cos\theta\bigg]\\
&- e_q\,\chi_1(s)\,\big[v_ev_q\,(1+\cos^2\theta)+2\,a_e\,a_q\,\cos\theta\big].
\end{align}
The vector and axial couplings of the electron to the $Z^0$ are, respectively, $v_e=4\,\sin^2\theta_w-1$ and $a_e=-1$. $\theta_w$ is the weak angle given by $\cos\theta_w=M_W/M_Z$, with $M_W$ ($\MZ$) being the mass of the $W$ ($Z^0$) boson. The vector coupling of the quarks to the $Z^0$ are $v_u=1-\frac{8}{3}\sin^2\theta_w$ and $v_d=v_s=-1+\frac{4}{3}\sin^2\theta_w$.  The axial couplings are instead $a_u=1$ and $a_d=a_s=-1$. The quark electric charge in units of the elementary charge is indicated by $e_q$. The functions $\chi_{1,2}(s)$ are given by~\cite{Chen:1994ar}
\begin{eqnarray}
    \chi_1(s) &=& \frac{1}{16\,\sin^2\theta_w\,\cos^2\theta_w}\,\frac{s\,(s-\MZ^2)}{(s-\MZ^2)^2+\GZ^2\,\MZ^2},\\
    \chi_2(s) &=& \frac{1}{256\,\sin^4\theta_w\,\cos^4\theta_w}\,\frac{s^2}{(s-\MZ^2)^2+\GZ^2\,\MZ^2},
\end{eqnarray}
and $\GZ$ is the $Z^0$ width.

From the expressions of the correlation coefficients in Eq.~(\ref{eq:Cqqbar}) it is clear that $\rho(\q,\qbar)$ in Eq.~(\ref{eq:rho}) depends on the quark flavour due to the weak coupling of quarks with $Z^0$. If the $Z^0$ contribution is neglected, the coefficients $\CUL$ and $\CMN$ vanish and $\CMM$ reduces to the quantity $\aNN(\theta)=\sin^2\theta/(1+\cos^2\theta)$. The latter coefficient correlates the transverse spin states of $\q$ and $\qbar$ if the $e^+e^-$ annihilation is purely mediated by the $\gamma^*$ (c.f. with Ref. ~\cite{Kerbizi:2023luv}).

%The quantity $\aNN(\theta)=\sin^2\theta/(1+\cos^2\theta)$ describes the correlation between the transverse spin states of $q$ and $\qbar$ originated by the tensor polarization of the $\gamma^*$. \red{Quark masses are vanishing. Rho is different for different quark flavours due to the weak coupling to the Z0. Neglecting the Z0 contribution, it reduces to the expression used in Refs. [Kerbizi-Artru, Kerbizi-Lonnblad-Martin].}

The spin density matrices of $q$ and $\qbar$ can be obtained from the joint spin density matrix as
\begin{align}\label{eq:rho q}
 \nonumber   \rho(q)&=\Tr_{\qbar}\,\rho(\q,\qbar)= \frac{1}{2}\,(\Iq+\Sq\cdot\boldsymbol{\sigma}_{q}),\\
    \rho(\qbar)&=\Tr_{\q}\,\rho(\q,\qbar)=\frac{1}{2}\,(\Iqbar+\Sqbar\cdot\boldsymbol{\sigma}_{\qbar}),
\end{align}
where the corresponding polarization vectors are obtained by inserting Eq.~(\ref{eq:rho}) in Eq.~(\ref{eq:rho q}) and they are
\begin{eqnarray}\label{eq:Sq,Sqbar initial}
\nonumber    \Sq&=&(\SqT,\SqL)=(0,0,-\CUL),\\
    \Sqbar&=&(\SqbarT,\SqbarL)=(0,0,\CUL).
\end{eqnarray}
The initial quarks are therefore longitudinally polarized due the parity-violating coupling with the $Z^0$. For the pure $\gamma^*$ contribution, the initial quarks are unpolarized. However, as described in Sec.~\ref{sec:wH}, the antiquark becomes transversely polarized if at least a hadron is emitted from the quark end of the string (and viceversa). %Rather, their spin states are correlated. %\red{This is true when the annihilation is mediated by a $\gamma^*$ or a $Z^0$, as can be seen from the more general expression of $\rho(\q,\qbar)$ including both $\gamma^*$ and $Z^0$ exchanges that can be found in Ref.~\cite{Chen:1994ar}. Parity violation due to the $Z^0$ contribution leads also to separate longitudinal polarizations for $\q$ and $\qbar$.}
%
%
%\end{subequations}
\begin{figure}[tbh]
\centering
\begin{minipage}[b]{0.45\textwidth}
\hspace{1.0em}
%\vspace{0.5em}
\includegraphics[width=0.8\textwidth]{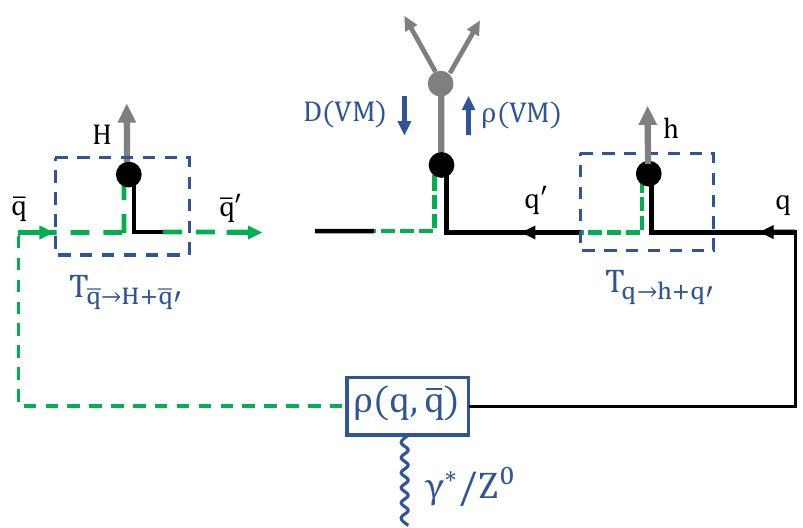}
\end{minipage}\vspace{-1em}
\caption{Representation of the spin-dependent fragmentation in \StringSpinner of the string stretched between $\q$ and $\qbar$.}
\label{fig:string fragmentation}
\end{figure}

\subsection{The recursive string fragmentation process}
 The string fragmentation process $q\,\qbar\rightarrow h,H,\dots$, where $h,H,\dots$ are the emitted hadrons, is simulated by \pythia as a recursive process of elementary quark splittings $q(k)\rightarrow h(p) + q'(k')$ and elementary antiquark splittings $\qbar(\kbar)\rightarrow H(P)+\qbarp(\kpbar)$. The quantity within the brackets after each particle indicates the corresponding four-momentum. This is shown schematically in Fig.~\ref{fig:string fragmentation}. The splittings are performed randomly with equal probability either from the $\q$ side or the $\qbar$ side of the string. Four-momentum conservation implies $k=p+k'$ and $\kbar=P+\kpbar$.
 
%In the splitting $q(k)\rightarrow h(p) +\qp(k')$ the fragmenting quark $q$ emits the hadron $h$ and leaves the recurring quark $q'$ to be fragmented in the next splitting $\qp\rightarrow h' +q''$ etc. The four-momenta of each particle are indicated in the brackets. Four-momentum conservation implies $k = p + k'$.
%Similarly, in the splittings $\qbar(\kbar)\rightarrow H(\pbar) +\qpbar(\kpbar)$ the fragmenting antiquark $\qbar$ emits the hadron $H$ and leaves the recurring quark $\qpbar$ to be fragmented in the next splitting $\qpbar\rightarrow H' +\qbar''$ etc. Four-momentum conservation implies $\kbar = \pbar + \kpbar$.

Important variables for the introduction of spin effects are the transverse momenta of the involved particles, defined with respect to the string axis (i.e. the $\zq$ axis in Fig.~\ref{fig:kinematics}). We indicate the transverse momenta of $\q(\qbar)$, $h(H)$ and $\qp (\qbarp)$ by $\kt(\ktbar)$, $\pt(\Pt)$ and $\kpt(\kptbar)$, respectively.
%
%as $\kt$, $\pt$ and $\kpt$, respectively.
They are related by \begin{align}\label{eq:kT conservation}
    \kpt = \kt-\pt, && \kptbar = \ktbar -\Pt,
\end{align} %The transverse momenta of $\qbar$, $H$ and $\qbarp$ are instead $\ktbar$, $\Pt$ and $\kptbar$, respectively, and they are related by $\kptbar = \ktbar-\Pt$. 
owing to momentum conservation.

%\subsection{Reweighting of hadrons in string fragmentation}\label{sec:reweight}
Before implementing the procedure that emulates the spin effects of the string+${}^3P_0$ model in \Pythia{}, \StringSpinner inspects each hadron ($h$, $H$, ...) emitted by \Pythia{} in string fragmentation and preliminary rejects the hadron if it is not a PM or a VM. %As described in the following, to enable spin effects in \Pythia{}, \StringSpinner applies a veto procedure on each hadron to emulate the string+${}^3P_0$ model in Ref.~\cite{Kerbizi:2023luv}.
If this selection is passed, the hadron is further rejected with a probability $1/2$ to keep the composition of the produced hadrons as in \Pythia{}. Then to emulate the spin effects the following reweighting procedure is applied to the hadrons that have passed the preliminary selections.

\subsubsection{Reweighting hadrons from the $\q$ end}
Let us suppose that the first hadron to be inspected is $h$, emitted from the quark end by the splitting $q\rightarrow h+\qp$. Then $h$ is accepted with probability \cite{Kerbizi:2024vpd}
\begin{eqnarray}\label{eq:wh} \wh(\kpt;\SqT)= \frac{1}{2}\,\left[1+c\,\frac{2\,\Im(\mu)}{|\mu|^2+\kptkpt}\,\SqT\cdot\left(\zq\times\kpt\right)\right],
\end{eqnarray}
which is the ratio between the probabilities for a polarized and an unpolarized splitting $\q\rightarrow h+\qp$ in the string+${}^3P_0$ model. It correlates the transverse momentum $\kpt$ of $\qp$ and the transverse polarization $\SqT$ of $\q$. For a non-vanishing $\SqT$, such correlation results in the Collins effect for $h$ due to Eq.~(\ref{eq:kT conservation}). The quark transverse polarization $\SqT$ is the component of $\textbf{S}_q=\Tr\,\sigmaq\,\rho(q)$ transverse to the string axis, where the spin density matrix $\rho(q)$ is obtained as in Eq.~(\ref{eq:rho q}). According to Eq.~(\ref{eq:Sq,Sqbar initial}), for the initial quark it is $\Sq=\textbf{0}$. Therefore the first hadron is emitted with a uniform azimuthal angle distribution.

The probability $\wh$ in Eq.~(\ref{eq:wh}) depends on the so-called complex mass $\mu=\Re(\mu)+\i\Im(\mu)$ and on the factor $c$. $\mu$ parametrizes the ${}^3P_0$ mechanism. $\Im(\mu)$ is related to transverse spin effects in hadronization whereas $\Im(\mu^2)=2\,\Re(\mu)\,\Im(\mu)$ is related to the longitudinal spin effects. The factor $c$ is $c=-1$ for a PM and $c=\fL$ for a VM, and it governs the relative sign and magnitude of the Collins effect for PM and VM emissions. The parameter $\fL=|\GL|^2/(2|\GT|^2+|\GL|^2)$ is the fraction of longitudinally polarized VMs. It is expressed in terms of the complex couplings $\GT$ and $\GL$ of quarks to VMs with transverse and longitudinal linear polarizations, respectively.

%It thus changes only the azimuthal distribution of $h$ produced by \Pythia{} to emulate the spin effects of the string+${}^3P_0$ model and, for a non-zero $\SqT$, it is responsible for the Collins effect in the emission of $h$. 

%Equation (\ref{eq:wh}) is the analogue of the probability introduced in Refs.~\cite{Kerbizi:2021StringSpinner, Kerbizi:2023cde} for the description of the spin effects in the DIS process.

%Unlike the procedure applied for DIS in Ref.~\cite{Kerbizi:2023cde}, hadrons emitted from the antiquark side are not rejected. which applies a veto procedure to emulate the rules of the string+${}^3P_0$ model inspired by the standalone implementation of the model in Ref. \cite{Kerbizi:2021M20}.

\subsubsection{Polarized decays}\label{sec:decay}
If the hadron $h$ is an unstable PM, its decay is handled by \Pythia{}. If $h$ is a VM, its decay is instead handled by \StringSpinner by taking into account the polarization of the meson encoded in the spin density matrix $\rho(h)$. In the rest frame of the meson, the density matrix reads \cite{Kerbizi:2023luv}
\begin{align}\label{eq:rho VM}
\rho_{\rm{aa'}}(h) = \frac{\Cqq_{\alpha 0}\rm\, \Trq\left[\Delta(\kpt)\,\Gammaq^a\,\sigmaq^{\alpha}\,\Gammaq^{a'\,\dagger}\,\Delta^{\dagger}(\kpt)\right]}{\Tr\left[ \Cqq_{\beta 0}\rm\, \Trq\left[\Delta(\kpt)\,\Gammaq^b\,\sigmaq^{\beta}\,\Gammaq^{b\,\dagger}\,\Delta^{\dagger}(\kpt)\right] \right]},
\end{align}
where $\Delta_{\qp}(\kpt)=\mu+\sigma_z\boldsymbol{\sigma}_{\rm T}\cdot \kpt$ is the ${}^3P_0$ propagator, $\Gamma^a=(\GT\sigma_x\sigma_z,\GT\sigma_y\sigma_z,\GL\Iden)$ is the coupling matrix of quarks to the VM and the correlation coefficients $\Cqq_{\alpha\,0}$ are obtained from Eq.~(\ref{eq:rho}). The spin density matrix of a VM depends on the coupling of quarks to the meson via the parameter $\fL$ and the parameter $\thetaLT=\arg(\GL/\GT)$. The latter describes the oblique polarization of the VM, i.e. the interference between longitudinal and transverse linear polarization states of the meson~\cite{Kerbizi:2021M20}. 

The angular distribution of the decay products in the rest frame of the VM is generated according to
\begin{eqnarray}
\frac{dN(p\rightarrow p_1, p_2,..)}{d\Phi(p_1,p_2,..)}&\propto& \hat{M}_a\,\rho_{aa'}\,\hat{M}^{\dagger}_{a'},
\end{eqnarray}
where $d\Phi(p_1,p_2,..)$ indicates the relevant differential phase space factor and $\hat{M}_{a}=\hat{M}_{a}(p\rightarrow p_1,p_2,..)$ is the amplitude describing the decay of a VM with linear polarization $a$ in the daughters $d_1,d_2,..$. The expression for $\hat{M}_a$ for the considered decay processes is given in Ref.~\cite{Kerbizi:2023luv}.

The decay of $h$ returns the decay matrix 
\begin{eqnarray}
D_{a'a}=\hat{M}^{\dagger}_{a'}\,\hat{M}_a,
\end{eqnarray}
required by the Collins-Knowles recipe \cite{Collins:1987cp,Knowles:1988vs} to propagate the information on the orientation of the decay hadrons to the leftover quark $\qp$, as schematically shown in Fig.~\ref{fig:string fragmentation}.

\subsubsection{The updated joint spin density matrix}
\label{sec:spin-propagation}
After the splitting $\q\rightarrow h+\qp$ has been performed, a new string piece stretched between $\qp$ and $\qbar$ remains to be fragmented. The spin state of the $\qp\qbar$ pair is described by the new joint spin density matrix $\rho(\qp,\qbar)$, which can be expressed as
\begin{equation}\label{eq:rho q'qb}
    \rho(\qp,\qbar)=\frac{1}{4}\,\C^{\qp\qbar}_{\alpha\beta}\,\sigmaqp^{\alpha}\otimes\sigmaqbar^{\beta}.
\end{equation}
Such matrix encodes the spin correlations between the leftover quark $\qp$ and the antiquark $\qbar$. The corresponding correlation coefficients are given by~\cite{Kerbizi:2023luv}
\begin{eqnarray}
    \C^{\qp\qbar}_{\alpha'\beta} = \C^{\q\qbar}_{\alpha\beta}\,M^{\qp}_{\alpha\alpha'}(\kpt)\times \left(\C^{\q\qbar}_{\sigma\,0}\,M^{\qp}_{\sigma\,0}(\kpt)\right)^{-1},
\end{eqnarray}
which are in general non-zero for all $\alpha'$ and $\beta$. The matrix elements $M^{\qp}_{\alpha\alpha'}(\kpt)$ depend on the spin of the hadron $h$ and they read~\cite{Kerbizi:2023luv}
\begin{equation}\label{eq:M matrices quark}
    \begin{aligned}
        M^{\q}_{\alpha\alpha'}  = \frac{1}{2}\times \begin{cases}
            \Tr\left[ \sigmaq^{\alpha'}\,\Delta(\kpt)\,\sigmaZq\,\sigmaq^{\alpha}\,\sigmaZq\,\Delta^{\dagger}(\kpt)\right] & h=\PM, \\
            \Tr\left[ \sigmaq^{\alpha'}\,\Delta(\kpt)\,\Gammaq^a\,\sigmaq^{\alpha}\,\Gammaq^{\dagger\,a'}\,\Delta^{\dagger}(\kpt)\right]\,D_{a'a} & h=\VM
        \end{cases}.
    \end{aligned}
\end{equation}

For a VM emission the decay matrix $D_{a'a}$ is required, while for a PM emission the indices $a$ and $a'$, and $D_{a'a}$ are removed.
The matrix $\rho(\qp,\qbar)$ now contains the information on the emission of $h$ from the quark end.

\subsubsection{Reweighting hadrons from the $\qbar$ end}\label{sec:wH}
The joint spin density matrix in Eq.~(\ref{eq:M matrices quark}) is used to evaluate the polarization vector of $\qbar$, which is given by $\textbf{S}_{\qbar}=\Tr\,\sigmaqbar\,\rho(\qp,\qbar)$. As shown in Ref.~\cite{Kerbizi:2023luv}, now $\SqbarT\neq \textbf{0}$ and $\qbar$ is transersely polarized [c.f. with Eq.~(\ref{eq:Sq,Sqbar initial})]. The hadron $H$ generated by \Pythia{} in the splitting $\qbar\rightarrow H+\qbarp$ is thus accepted with probability~\cite{Kerbizi:2024vpd}
\begin{eqnarray}\label{eq:wH}
w_H(\kptbar;\SqbarT)= \frac{1}{2}\,\left[1+c\,\frac{2\,\Im(\mu)}{|\mu|^2+\kptbarkptbar}\,\SqbarT\cdot\left(\zqbar\times\kptbar\right)\right],
\end{eqnarray}
%where $\SqbarT$ is the component of $\textbf{S}_{\qbar}$ transverse to the string axis.
%Likewise to $w_h$ in Eq.~(\ref{eq:wh}), $w_H$ is the ratio between the probabilities for a polarized splitting and an unpolarized splitting according to the string+${}^3P_0$ model.
Since $\rho(\qp,\qbar)$ depends on the transverse momentum $\pt$ of $h$, Eq.~(\ref{eq:wH}) leads to correlations between the azimuthal angles of $\pt$ and $\Pt$.
$\wH$ can thus be interpreted as the conditional probability for emitting $H$ from the $\qbar$ side of the string if the hadron $h$ has been emitted from the $\q$ side. This is the mechanism for the generation of the Collins asymmetry for hadrons produced back-to-back in $e^+e^-$ in the string+${}^3P_0$ model.

For the decay of $H$ and the propagation of the spin information after its generation, the updated joint spin density matrix $\rho(\qp,\qbar)$ is used. The explicit steps can be found in Ref.~\cite{Kerbizi:2023luv}. 

\subsection{Termination of string fragmentation}
The emission of hadrons by elementary splittings from the quark side and the antiquark sides of the string is continued recursively until the exit condition of the string fragmentation process is called by \Pythia{}. The remaining string piece stretched between a quark $\q_m$ and an antiquark $\qbar_n$ must be fragmented by one last breaking where a $\q'\,\qbar'$ pair is produced and thus followed by the formation of the final two hadrons $h=\q_m\qbar'$ and $H=\q'\qbar_n$. 

If the splitting previous to the call for the exit condition was taken from the
antiquark side, we consider $h$ to be produced by the splitting
$\q_m\rightarrow h+\q'$ and project the $\q'\qbar_n$
state onto the hadronic state $H$. The hadron $h$ is thus accepted by the probability in Eq.~(\ref{eq:wh}) and decayed as in Sec.~\ref{sec:decay}. The hadron $H$ is taken to be unpolarized. If $h$ is not accepted by this procedure, \Pythia{} rejects both hadrons $h$ and $H$ and restarts the generation procedure of the final two hadrons only.

If the previous splitting was taken from the quark side, the hadron $H$ is considered to be produced by the splitting $\qbar\rightarrow H+\qbarp$, reweighted according to Eq.~(\ref{eq:wH}) and decayed with a similar procedure to Sec.~\ref{sec:decay}. In this case, the hadron $h$ is taken to be unpolarized.

%This recipe is somewhat simplified with respect to that proposed in Ref.~\cite{Kerbizi:2023luv}. We checked, however, that the simulation results obtained with the two recipes do not differ to a noticeable degree at the considered \CMS energy.
%This is due to the fact that the spin information decays along the fragmentation chain, and the possible spin effects in the production of the final two hadrons are negligible.

\section{Program files and instructions}\label{sec:program files}
The main file of the \StringSpinner package is \texttt{StringSpinner.h}, which contains the implementation of the \texttt{UserHooks} class for the introduction of the spin effects in \pythia{}. It has been updated 
as compared to Ref. \cite{Kerbizi:2023cde} by including the modifications that allow to activate quark spin effects for $e^+e^-$ annihilation in \Pythia. In particular, it requires the header file \texttt{CorrelationCoefficients.h}, which is a new file. This file implements the \texttt{CorrelationCoefficients} class that represents the joint spin density matrix of the initial $\q\qbar$ pair in Eq.~(\ref{eq:rho}) as well as the methods for the evaluation of the updated density matrix after the emission of each hadron [see, e.g., Eq.~(\ref{eq:rho q'qb})]. To evaluate the updated density matrix, the \texttt{CorrelationCoefficients} class makes use of new \texttt{Fortran} routines that have been included in the \texttt{mc3P0.f90} file (already present in the previous version of \StringSpinner). Such routines implement the actual operations with Pauli matrices appearing in Eq.~(\ref{eq:rho VM}) and Eq.~(\ref{eq:M matrices quark}).

To generate annihilation events with spin effects we have included the two new main programs \texttt{ee.cc} and \texttt{eeToqq.cc}. The former allows to generate complete $e^+e^-$ annihilation events, where the produced $\q\qbar$ pair is described by the joint spin density matrix in Eq.~(\ref{eq:rho}). The latter allows to simulate the hadronization of a $\q\qbar$ pair with full freedom of choosing quark flavours, their momenta as well as their joint spin density matrix.

The remaining files of the package have not been changed and their description can be found in Ref.~\cite{Kerbizi:2023cde}.

The complete \StringSpinner package can be downloaded from
\texttt{gitlab}\footnote{\href{https://gitlab.com/albikerbizi/stringspinner.git}{https://gitlab.com/albikerbizi/stringspinner.git}.}. It also
includes a configure script and a \texttt{Makefile} for the
compilation of the main program, which will be explained in more detail
in Sec. \ref{sec:execution}.

\subsection{The main program}
The example main program \texttt{ee.cc} allows to simulate $e^+e^-$ annihilation events at the \CMS{} energy $\sqrt{s}=10.6\,\GeV$, as in the BELLE and BABAR experiments. The structure of the main program is the one of standard \pythia with few additions, summarized below.

The schematic structure of the main program is
\small
\begin{verbatim}
#include "Pythia8/Pythia.h"
#include "StringSpinner.h"
using namespace Pythia8;
int main() {
  Pythia pythia;
  Event& event = pythia.event;
  auto fhooks = 
      std::make_shared<SimpleStringSpinner>();
  fhooks->plugInto(pythia);

  // Choice of beams.
  double eElec  = 8.0;
  double ePos	  = 3.5;
  pythia.readString("Beams:frameType = 2");
  pythia.readString("Beams:idA =  11");
  pythia.readString("Beams:idB = -11");
  pythia.settings.parm("Beams:eA", eElec);
  pythia.settings.parm("Beams:eB", ePos);
  
  // Standard Pythia settings.
  pythia.readString("WeakZ0:gmZmode = 1");    
  pythia.readString("WeakSingleBoson:ffbar2gmZ = on");
  pythia.readString("23:onMode = off");         
  pythia.readString("23:onIfAny = 1 2 3");
  // ....

  // Switch off parton showers in Pythia.
  pythia.readString("PartonLevel:FSRinResonances = off");
  pythia.readString("PartonLevel:FSR = off");
  pythia.readString("PartonLevel:ISR = off");

  // StringSpinner settings.
  // ...

  if ( !pythia.init() ) exit(-1);

  for(int iEvent=0;iEvent<nEvents;iEvent++){
    if( !pythia.next() ) continue;
    // Boost event to the cms.
    // ...
    // Analysis of the Pythia event record using
    // the standard tools.
    // ...
  }
  // ...
  return 0;
}
\end{verbatim}
\normalsize The \texttt{StringSpinner.h} file must be included in the
main program. To activate the spin effects, the creation of the
\setting{Pythia} object by \setting{Pythia pythia}, must be followed by
the creation of a shared pointer to a \setting{SimpleStringSpinner}
object. The pointer is named \setting{fhooks} and it is
created by \setting{auto fhooks =
  std::make\_shared<SimpleStringSpinner>()}. The pointer to the
created \setting{UsersHooks} object is passed to the \setting{Pythia} object by \setting{fhooks->plugInto(pythia)}. This
allows \pythia to simulate the spin effects for the processes that can
be handled. It also allows the user to communicate with \StringSpinner using the \pythia settings system.

%As described in Ref. \cite{Kerbizi:2021StringSpinner},
%\setting{SimpleStringSpinner} is the ad hoc implementation of the
%\setting{UserHooks} class for the insertion of spin effects in
%\pythia.

To change the settings of the generator, the standard \pythia
command \setting{pythia.readString()} can be used before the
\setting{pythia.init()} command. The same command can be used also for
the settings of \StringSpinner. The \setting{pythia.readString()}
command takes a character string of the form \setting{"Parameter =
  value"} as argument. To set the free parameters of the string+${}^3P_0$
model in \StringSpinner the following settings can be used:
\begin{itemize}\itemsep -1mm
\item \settingval{StringSpinner:re(Mu)}{$\Re(\mu)$}
\item \settingval{StringSpinner:im(Mu)}{$\Im(\mu)$}
\item \settingval{StringSpinner:GLGT}{$|\GL/\GT|$}
\item \settingval{StringSpinner:thetaLT}{$\thetaLT$}.
\end{itemize}
%\textit{reMu} and \textit{imMu} are the values of the real and
%imaginary parts of $\mu$ respectively. \textit{gLgT} indicates the value of the coupling constant ratio $|G_{\rm L}/G_{\rm T}|$ entering the parameter $\fL$, whereas \textit{thetaLT} is the parameter describing the oblique polarization of the VMs. The parameter $\fL$ can take values between $0$ and $1$, whereas $\thetaLT$ is expected to take values between $-\pi$ and $\pi$.
% The complementary methods \texttt{fhooks.getReMu()}, \texttt{fhooks.getImMu()}, \texttt{fhooks.getFL()} and \texttt{fhooks.getThetaLT()} allow instead to access the free parameters. Each method returns a \texttt{double} variable with the value of the corresponding parameter.

The initialization of the joint spin density matrix of the produced $\q\qbar$ pair can be modified with the standard \Pythia{} setting
\begin{itemize}\itemsep -1mm
\item \texttt{WeakZ0:gmZmode}
\end{itemize}
where \texttt{gmZmode} can take the values \texttt{0} (full interference between $\gamma^*$ and $Z^0$, as in Eq.~(\ref{eq:rho})), \texttt{1} (only $\gamma^*$ contribution) and \texttt{2} (only $Z^0$) contribution. The default setting is \texttt{mgZmode=0}.

Alternatively, the joint spin density matrix of the fragmenting $\q\qbar$ pair can be initialized with the commands
\begin{itemize}\itemsep -1mm
\item \texttt{StringSpinner:spinCorrCoeff0j}
\item \texttt{StringSpinner:spinCorrCoeffj0}
\item \texttt{StringSpinner:spinCorrCoeffxj}
\item \texttt{StringSpinner:spinCorrCoeffyj}
\item \texttt{StringSpinner:spinCorrCoeffzj}
\end{itemize}
where each parameter is expected to take as input a vector giving the values of the coefficients. If any of these commands is used, \StringSpinner sets the corresponding correlation coefficients to the input values and all other coefficients equal to zero, except $\Cqq_{00}$, which is always set to $1/4$. For instance
\begin{itemize}\itemsep -1mm
\item \texttt{StringSpinner:spinCorrCoeffj0 = 0.0,1.0,0.0}
\end{itemize}
performs the setting $\C^{\q\qbar}_{x0}=0.0$, $\C^{\q\qbar}_{y0}=1.0$ and $\C^{\q\qbar}_{z0}=0.0$. It corresponds to a string stretched between a quark with polarization vector $\Sq=(0,1,0)$, i.e. transverse polarization along the $\yq$ axis, and an unpolarized antiquark.

\subsection{Installation and running}\label{sec:execution}
Assuming that \pythia~8.3 has already been installed, the \StringSpinner package can be downloaded and installed using the following shell commands
\small
\begin{verbatim}
git clone https://gitlab.com/albikerbizi/stringspinner
cd stringspinner
./configure path/to/pythia/installation/directory
make
\end{verbatim}
\normalsize This produces the file \texttt{Makefile.inc}, which stores
the path to the installation directory of \pythia and is included in
the file \texttt{Makefile}. The latter gathers the necessary commands
to compile the main program. The main program can then be compiled together with the other files of
the package using the command \texttt{make ee}, which produces the
executable \texttt{ee}, and run with \texttt{./ee}. The cancellation
of the files generated by the compilation procedure can be done using
the \texttt{make clean} command.

\subsection{Setting the free parameters}\label{sec:validation}
The new implementation of the spin effects in the \StringSpinner package has been validated by comparing the resulting kinematical distributions of the final state hadrons and the transverse spin asymmetries with those obtained with the previous version of \StringSpinner~\cite{Kerbizi:2023cde}. For the comparison, we have simulated fragmentations of $u\bar{u}$ strings with the present package and fragmentations of $u(ud)_0$ strings with the previous package, taking in both cases the initial $u$-quark to be transversely polarized. For the $u\bar{u}$ string, the $\bar{u}$ is taken unpolarized. This allowed to set up similar conditions in the two versions of the generator. The results obtained with the two generators were similar for different settings of the free parameters of the string+${}^3P_0$ model, thereby validating the new \StringSpinner package presented in this work.

%The default settings of the free parameters of th for this version of \StringSpinner we have performed a systematic tuning of the free parameters $\Re\mu$, $\Im\mu$, $\fL$ and $\thetaLT$ of the string+${}^3P_0$ model using the \texttt{Professor} package. %The spin-independent parameters have been tuned to the $e^+e^-$ cross-section data measured by BELLE as a function of the invariant mass of hadron pairs produced in the same hemisphere of the event and in bins of the fractional energy of the pair~[]. Only the oppositely charged pairs $\pi^+\pi^-$, $\pi^+K^-$ and $K^+K^-$ have been used. We have obtained
%\begin{verbatim}
%    StringZ:aLund=xx
%    StringZ:bLund=xx
%    StringFlav:mesonUDvector=xx.
%\end{verbatim}

Concerning the free parameters of the string+${}^3P_0$ model, in the present paper we use the setting
\begin{itemize}\itemsep -1mm
\item \verb|StringSpinner:re(Mu)  = 0.11|
\item \verb|StringSpinner:im(Mu)  = 0.33|
\item \verb|StringSpinner:GLGT    = 3.11|
\item \verb|StringSpinner:thetaLT = 0.09|.
\end{itemize}
%\red{In particular, the value used for $|\GL/\GT|$ gives $\fL=0.83$, meaning that VMs are produced in string fragmentation mainly with longitudinal linear polarization (i.e. along the string axis). The small but non-zero value obtained for $\thetaLT$ allows for interferences between the longitudinal and the transverse polarizations of the VMs.}

These values of the free parameters have been obtained by a preliminary tuning of \StringSpinner using the \texttt{Professor} package~\cite{Buckley:2009bj}. The latter allows to build surrogate functions that parametrize the relevant \StringSpinner output as a function of the free parameters $\Re(\mu)$, $\Im(\mu)$, $|\GL/\GT|$ and $\thetaLT$ using polynomials. The tune parameters are found by a $\chi^2$ minimization procedure that makes use of the parametrized \StringSpinner results on transverse-spin asymmetries and the data on Collins and dihadron asymmetries in SIDIS from COMPASS~\cite{COMPASS:2014bze,COMPASS:2014ysd} and HERMES~\cite{HERMES:2010mmo}, and the Collins asymmetries in $e^+e^-$ reactions from BELLE~\cite{Belle:2008fdv,Belle:2019nve} and BABAR~\cite{BaBar:2013jdt,BaBar:2015mcn}.

The complete description of the tuning of \StringSpinner is left for a separate work~\cite{tuning}.

\section{Example: polar angle dependence of the Collins asymmetry $\AOneTwo$}\label{sec:results}
As an example of usage of the present \StringSpinner package we consider the calculation of the Collins asymmetry $\AOneTwo$ following the experimental analyses by BELLE~\cite{Belle:2008fdv} and BABAR~\cite{BaBar:2013jdt}. The asymmetry $\AOneTwo$ is measured by looking at the correlation between the azimuthal angle $\phi_1$ of the hadron $h_1$ and $\phi_2$ of hadron $h_2$, where $h_1$ and $h_2$ are produced almost back-to-back in the \CMS\ of the $e^+e^-$ event. The azimuthal angles are measured with respect to the plane formed by the thrust axis $\n$ and the momentum $\pmin$ of the beam electron. Only events characterized by a high value of the thrust variable $T$ are selected by requiring $T>0.8$. In each event, pairs of back-to-back hadrons are formed by requiring $(\Pbf_1\cdot\n)(\Pbf_2\cdot\n)<0$. The association of hadrons to the wrong hemisphere is minimized by the selection $Q_{\rm T}>3.5\,\GeV/c$, where $\QT$ is the transverse momentum of the virtual photon in the rest frame of the hadron pair.

The angular distribution of the back-to-back hadrons can be written as ~\cite{Boer:2008fr}
\begin{eqnarray}\label{eq:N12}
N_{12}(z_1,\PTa,z_2,\PTb,\cos\theta;\phi_{12})\propto 1 +  \AOneTwo\,\cos(\phi_{12}),
\end{eqnarray}
where $\phi_{12}=\phi_1+\phi_2$. The variables $z_{1(2)}$ and $P_{1(2)\,\rm T}$ indicate, respectively, the fractional energy and the transverse momentum with respect to the thrust axis of hadron $h_{1(2)}$.

The partonic expression of the Collins asymmetry $\AOneTwo$ can be written as~\cite{Boer:2008fr}~\footnote{Note that in this work the angular factor $\sin^2\theta/(1+\cos^2\theta)$ is included in the partonic expression of the asymmetry, like in experimental analyses and at variance with Ref.~\cite{Kerbizi:2024vpd}.}
\begin{eqnarray}\label{eq:A12}
    \AOneTwo &=& \frac{\sin^2\theta}{1+\cos^2\theta}\\
\nonumber    &\times&\frac{\sum_{q,\qbar}\,e_q^2\,\frac{\PTa}{z_1\,m_{h_1}}\HqhOne(z_1,\PTa)\,\frac{\PTb}{z_2\,m_{h_2}}\HqhTwo(z_2,\PTb)}{\sum_{q,\qbar}\,e_q^2\,\DOne(z_1,\PTa)\,\DTwo(z_2,\PTb)},
\end{eqnarray}
where $e_q$ is the charge of the quark $q$ in units of the elementary charge and $m_{h_{1(2)}}$ indicates the mass of the hadron $h_{1(2)}$. The function $H_{1q}^{\perp\,h}$ indicates the Collins fragmentation function (FF) describing the Collins effect, namely the fragmentation of a transversely polarized quark $q$ in the unpolarized hadron $h$. $D_{1q}^h$ is instead the spin-averaged FF describing the fragmentation of an unpolarized quark $q$ in the unpolarized hadron $h$. The summation over both quark and antiquark indices in Eq.~(\ref{eq:A12}) indicates the fact that each hadron can be associated either with the quark or with the antiquark fragmentations.

To minimize systematic uncertainties, the experimental analyses make use of the normalized yield $R_{12}(\phi_{12})=N_{12}(\phi_{12})/\langle N_{12}\rangle$ in a chosen kinematic bin, where $\langle N_{12}\rangle$ is the average yield in that bin. The ratios $R_{12}^{U}$, $R_{12}^L$ and $R_{12}^C$ are constructed for pairs with unlike charge ($U$), like charge ($L$) and for any pair of charged ($C$) hadrons. The Collins asymmetry is finally extracted from the ratio $R_{12}^{UL(UC)}$, whose angular dependence reads [cf. Eq.~(\ref{eq:N12})]
\begin{eqnarray}
    R_{12}^{UL(UC)}=\frac{R_{12}^U}{R_{12}^{L(C)}}\simeq 1+ A_{12}^{UL(UC)}\,\cos\phi_{12}.
\end{eqnarray}

The measured Collins asymmetry is thus $\AOneTwo^{UL(UC)}\simeq \AOneTwo^{U}-\AOneTwo^{L(C)}$. It is afterwards rescaled by a coefficient that takes into account the dilution of the asymmetry induced by the mismatch between the thrust axis $\n$ and the true $\q\qbar$ axis in the \CMS~\cite{Belle:2008fdv,BaBar:2013jdt}. A further rescaling that account for the dilution caused by the fragmentation of charm quarks is performed.

\begin{figure}[th]
\centering
\begin{minipage}[b]{0.50\textwidth}
%\hspace{-2.0em}
\includegraphics[width=1.0\textwidth]{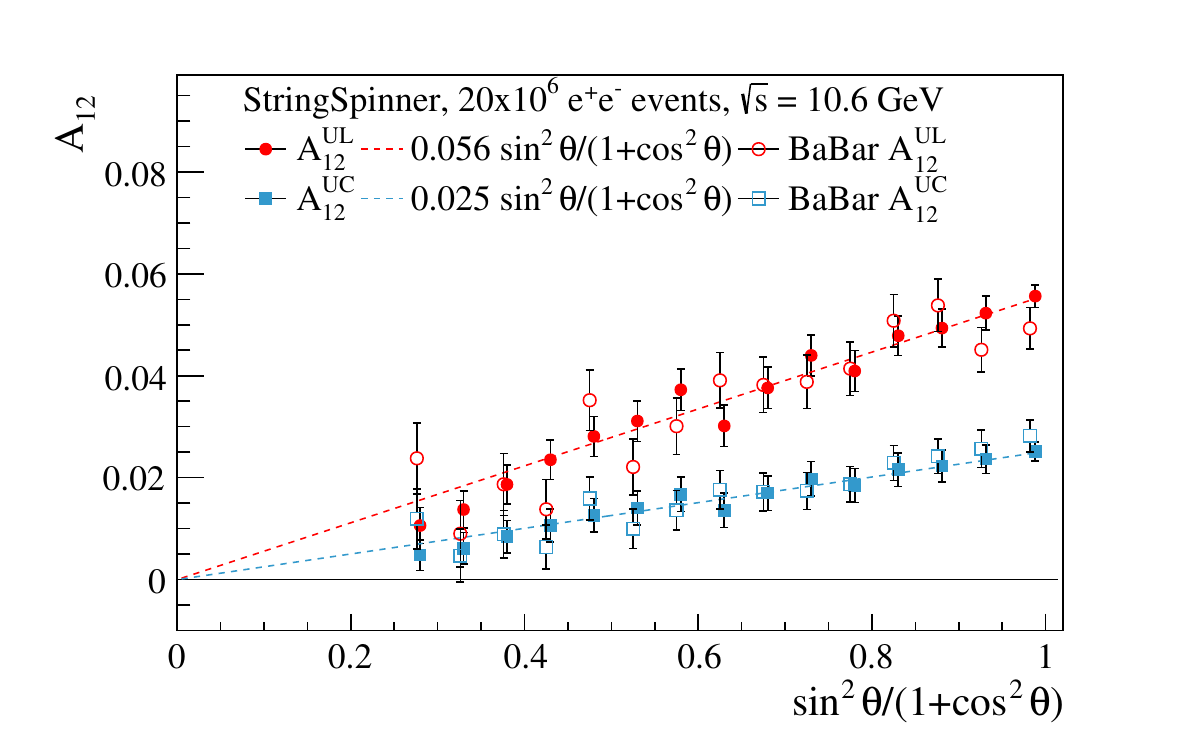}
%\vspace{1.8em}
\end{minipage}
\caption{Simulated $\AOneTwoUL$ (full circles) and $\AOneTwoUC$ (full squares) asymmetries as a function of $\sin^2\theta/(1+\cos^2\theta)$ obtained from $20\,10^6$ $e^+e^-$ events at $\sqrt{s}=10.6\,\GeV$. The straight lines represent the linear fits to the asymmetries. The corresponding asymmetries measured by BABAR~\cite{BaBar:2013jdt} are given by the open points.}
\label{fig:A12-sim}
\end{figure}

To obtain the simulated Collins asymmetry we use the true $\q\qbar$ axis for the evaluation of the azimuthal angle $\phi_{12}$ and allow only for the production of quarks $q=u,d,s$. %The dependence of the simulated $\AOneTwo$ asymmetry on the fractional energies as well as the transverse momenta of the produced hadrons has been studied in detail in Ref.~\cite{Kerbizi:2023luv}.
In this work we focus on the dependence of the Collins asymmetry on the angular factor $\sin^2\theta/(1+\cos^2\theta)$, predicted to be linear from Eq.~(\ref{eq:A12}) and observed experimentally~\cite{Belle:2008fdv,BaBar:2013jdt}. The simulated Collins asymmetry as a function of $\sin^2\theta/(1+\cos^2\theta)$ in the kinematic configuration of BABAR~\cite{BaBar:2013jdt} is shown in Fig.~\ref{fig:A12-sim} for the $UL$ (full circles) and $UC$ (full squares) charge configurations. The straight lines are the results of linear fits to the $\AOneTwoUL$ and $\AOneTwoUC$ asymmetries. Likewise to the experimental analysis we have applied the selection $z_{1(2)}>0.1$. As can be seen, the dependence of each simulated asymmetry on $\sin^2\theta/(1+\cos^2\theta)$ is very well described by a straight line with vanishing intercept, demonstrating the correctness of the implementation of the quark spin effect for the $e^+e^-$ process. Moreover, the $\AOneTwoUL$ and $\AOneTwoUC$ asymmetries are in excellent agreement with the corresponding asymmetries measured by BABAR~\cite{BaBar:2013jdt} (open points). The slopes of the straight lines in Fig.~\ref{fig:A12-sim} also agree with the corresponding measurements performed by BELLE~\cite{Belle:2008fdv}. %These comparisons confirm the correctness of the tuned values of the free parameters.

A more detailed study of the \StringSpinner predictions will be given in Ref.~\cite{tuning}.

\section{Conclusions}\label{sec:conclusions}
We have extended the \StringSpinner package to enable the simulation of $e^+e^-$ annihilation with quark spin effects in the \Pythia{} generator by using the string+${}^3P_0$ model~\cite{Kerbizi:2023luv}. The new package allows to activate spin effects in \Pythia{} for both DIS and $e^+e^-$. The $e^+e^-$ annihilation is considered at leading order and the final state parton shower is switched off. The $e^-$ and $e^+$ beams are taken to be unpolarized. The $\q\qbar$ produced in the $e^+e^-$ annihilation is taken to have correlated spin states described by a joint spin density matrix $\rho(\q,\qbar)$, whose implemented expression takes into account both annihilation channels arising via a $\gamma^*$ or a $Z^0$ boson. Alternatively, to enable the study of the spin effects predicted by the string+${}^3P_0$ model we allow for the possibility of a user-defined joint spin density matrix.

In this work we use a setting for the free parameters of the string+${}^3P_0$ model obtained from an ongoing work on the tuning of \StringSpinner using transverse-spin asymmetry data from the $e^+e^-$ and SIDIS reactions~\cite{tuning}. %The present version of \StringSpinner has been tuned to SIDIS and $e^+e^-$ data on transverse-spin effects~\cite{tuning} and it allows to reproduce the spin effects in both reactions.
We evaluate as an example the Collins asymmetry $\AOneTwo$ as a function of the angular factor $\sin^2\theta/(1+\cos^2\theta)$ obtaining an excellent agreement with the data. %The simulated asymmetry is moreover found to be in excellent agreement with the data from the BELLE and BABAR experiments confirming the correctness of the tuned free parameters of the string+${}^3P_0$ model.

Finally, the work presented in this paper provides the framework for further advancements in the simulation of spin effects in hadronization. Examples are the application of the string+${}^3P_0$ model to the simulation of the fragmentation of generic string configurations produced by the final state parton shower and a more complete simulation of spin effects in the target fragmentation region of DIS events by using the recent extension of the string+${}^3P_0$ model to baryon production~\cite{Kerbizi:2025keh}.

\section*{Declaration of competing interest}
The authors declare that they have no known competing financial interests or personal relationships that could have appeared to
influence the work reported in this paper.

\section*{Acknowledgement}
This work was done in the context of the project “SPINFRAG: Spin-dependent string fragmentation”, funded by the European Union under Marie Skłodowska-Curie Actions (MSCA), grant agreement ID 101107452.

%% References


\bibliographystyle{elsarticle-num}

\bibliography{bibliography.bib}






% \end{thebibliography}

\end{document}